\documentclass[runningheads]{llncs}
\usepackage[T1]{fontenc}
\usepackage{graphicx}
\usepackage{booktabs}
\usepackage{algorithm}
\usepackage{algpseudocode}
\usepackage{booktabs}

\usepackage[numbers]{natbib}  
\usepackage[colorlinks=true, linkcolor=black, citecolor=blue]{hyperref}
\usepackage{multirow}
\usepackage{makecell} 
\usepackage{amsmath}
\usepackage{amssymb}
\usepackage{textcomp}
\usepackage{graphicx} 
\usepackage{braket} 
\usepackage{rotating}
\usepackage{changepage} 
\begin{document}
\title{Physics-informed Evolution: An Evolutionary Framework for Solving Quantum Control Problems Involving the Schr\"{o}dinger Equation}
\titlerunning{Physics-informed Evolution}
\author{Kaichen Ouyang$^{*}$\inst{1}\orcidID{0009-0003-5937-5229} \and
Mingyang Yu$^{*}$\inst{2}\orcidID{0009-0000-5903-6014} \and
Zong Ke\inst{3}\orcidID{0009-0006-0450-5614} \and
Jun Zhang\inst{2, 5}\orcidID{0000-0001-7835-9871}\and 
Yi Chen$^{\dagger}$\inst{4}\orcidID{0000-0002-7027-6542} \and
Huiling Chen$^{\dagger}$\inst{4}\orcidID{0000-0002-7714-9693}}
\authorrunning{K. Ouyang et al.}
\institute{
Department of Physics, University of Science and Technology of China, Hefei 230026, China
\email{oykc@mail.ustc.edu.cn}
\and
College of Artificial Intelligence, Nankai University, Tianjin 300350, China
\email{1120240312@mail.nankai.edu.cn}
\and
Faculty of Science, National University of Singapore, Singapore
\email{a0129009@u.nus.edu}
\and
Department of Computer Science, Wenzhou University, Wenzhou 325035, China
\email{kenyoncy2016@gmail.com, chenhuiling.jlu@gmail.com}
\and
Hanyang University ERICA Campus, South Korea
\email{junzhang@ieee.org}
\\
$^{*}$Equal contribution. $^{\dagger}$Corresponding authors.
}

%
\maketitle              
\begin{abstract}
Physics-informed Neural Networks (PINNs) show that embedding physical laws directly into the learning objective can significantly enhance the efficiency and physical consistency of neural network solutions. Similar to optimizing loss functions in machine learning, evolutionary algorithms iteratively optimize objective functions by simulating natural selection processes. Inspired by this principle, we ask a natural question: can physical information be similarly embedded into the fitness function of evolutionary algorithms? In this work, we propose Physics-informed Evolution (PIE), a novel framework that incorporates physical information derived from governing physical laws into the evolutionary fitness landscape, thereby extending Physics-informed artificial intelligence methods from machine learning to the broader domain of evolutionary computation. As a concrete instantiation, we apply PIE to quantum control problems governed by the Schr\"{o}dinger equation, where the goal is to find optimal control fields that drive quantum systems from initial states to desired target states. We validate PIE on three representative quantum control benchmarks: state preparation in V-type three-level systems, entangled state generation in superconducting quantum circuits, and two-atom cavity QED systems. Within the PIE framework, we systematically compare the performance of ten single-objective and five multi-objective evolutionary algorithms. Experimental results demonstrate that by embedding physical information into the fitness function, PIE effectively guides evolutionary search, yielding control fields  with high fidelity, low state deviation, and robust performance across different scenarios. Our findings further suggest that the Physics-informed principle extends naturally beyond neural network training to the broader domain of evolutionary computation.

\keywords{Physics-informed Neural Networks \and Evolutionary Algorithms \and Quantum Control \and Schr\"{o}dinger Equation \and Physics-informed Evolution}
\end{abstract}
\section{Introduction}

Physics-informed Neural Networks (PINNs) have achieved remarkable success across diverse domains, including fluid mechanics, heat transfer, and inverse problems\cite{raissi2019physics}. Their core idea is to embed physical laws derived from governing equations directly into the neural network loss function, enabling the model to simultaneously fit observational data and satisfy underlying physics. This paradigm offers several key advantages: it alleviates the reliance on large labeled datasets, enforces physical consistency even in data-sparse regions, and provides a unified framework for solving both forward and inverse problems, where forward problems refer to predicting system behavior from known physical laws and parameters, and inverse problems refer to inferring unknown parameters or physical laws from observational data. As a result, PINNs have been successfully applied to complex scenarios such as turbulence modeling, thermal management, and parameter identification in partial differential equations, demonstrating their versatility and effectiveness in integrating domain knowledge with deep learning\cite{karniadakis2021physics}.

From an optimization perspective, neural network training can be viewed as solving a non-convex optimization problem, typically addressed by gradient-based optimizers such as SGD and Adam\cite{bottou2018optimization,kingma2014adam}. In parallel, evolutionary algorithms (EAs) represent a distinct family of black-box optimizers that simulate natural selection through mutation, crossover, and selection operators, enabling them to iteratively optimize objective functions without requiring gradient information\cite{ouyang2026learn}.Numerous studies have employed evolutionary algorithms as substitutes for gradient-based methods in training neural networks, a field also known as neuroevolution, with successful applications across various domains such as reinforcement learning, neural architecture search, and hyperparameter optimization\cite{zhou2021survey,liu2021survey,li2024bridging}. Notably, some studies have shown that optimizers like SGD can, under certain conditions, be viewed as evolutionary strategies with extremely low mutation rates, revealing a deep connection between evolutionary search and gradient-based learning\cite{whitelam2021correspondence}. In recent years, with the widespread adoption of large language models, research has also explored the use of evolutionary algorithms for fine-tuning Large Language Models (LLMs)\cite{qiu2025evolution}.

Specifically regarding the training process of PINNs, from an optimization perspective, it can be viewed as a multi-objective non-convex optimization problem, where the objectives typically consist of multiple components such as data observation loss, partial differential equation residual loss, boundary condition loss, and initial condition loss\cite{wang2021understanding}. Considering that gradient-based methods dominate neural network training, while neuroevolution\cite{stanley2019designing}, as a gradient-free alternative, shares a profound connection with gradient-based methods, a natural question arises: given the success of PINNs in incorporating physical knowledge into gradient-based optimization, can the same principle be extended to evolutionary computation. In other words, if physical information can guide the loss landscape in neural network training, can it similarly shape the fitness landscape to guide population-based search? Inspired by this, we propose Physics-informed Evolution (PIE), a framework that embeds physical information derived from governing equations directly into the fitness function of evolutionary algorithms. Within this framework, physical information is encoded as the fitness function, guiding the iterative update of populations in evolutionary algorithms.

Considering that PINNs are fundamentally about leveraging physical information to better learn governing equations and thus align with real physical laws, analogously, as a concrete instantiation, we apply PIE to quantum control problems governed by the Schr\"{o}dinger equation, where the goal is to find optimal time-dependent control fields that steer quantum systems from initial states to desired target states. Such problems are characterized by high nonlinearity, where the evolution of quantum states is highly sensitive to control fields, and often involve multiple local optima, making it easy for traditional optimization methods to become trapped in suboptimal control strategies. We validate PIE on three representative quantum control benchmarks: state preparation in V-type three-level systems, entangled state generation in superconducting quantum circuits, and two-atom cavity QED systems\cite{dong2015sampling}. Across these scenarios, we systematically evaluate ten single-objective and five multi-objective evolutionary algorithms within the PIE framework.
The main contributions are as follows:
\begin{itemize}
	\item PIE is proposed as a general framework that extends the physics-informed principle from gradient-based learning to evolutionary computation by embedding physical laws into fitness functions.
	\item Both single-objective and multi-objective variants of PIE are developed, and multiple evolutionary algorithms are systematically evaluated within this framework.
	\item Comprehensive empirical validation is provided on three quantum control benchmarks, identifying effective algorithm choices for each specific scenario and demonstrating that PIE consistently guides evolutionary search toward high-fidelity, energy-efficient control fields.
\end{itemize}

The remainder of this paper is organized as follows. Section 2 reviews related work on PINNs, EAs, and the intersection of learning and evolution. Section 3 introduces the three quantum control problems used as benchmarks. Section 4 presents the PIE framework, including search space encoding and the construction of physics-informed fitness functions. Section 5 reports experimental results and analysis. Section 6 concludes the paper and discusses future directions.

\section{Related Works}

\begin{figure}[t]
	\centering
	\includegraphics[width=0.75\textwidth]{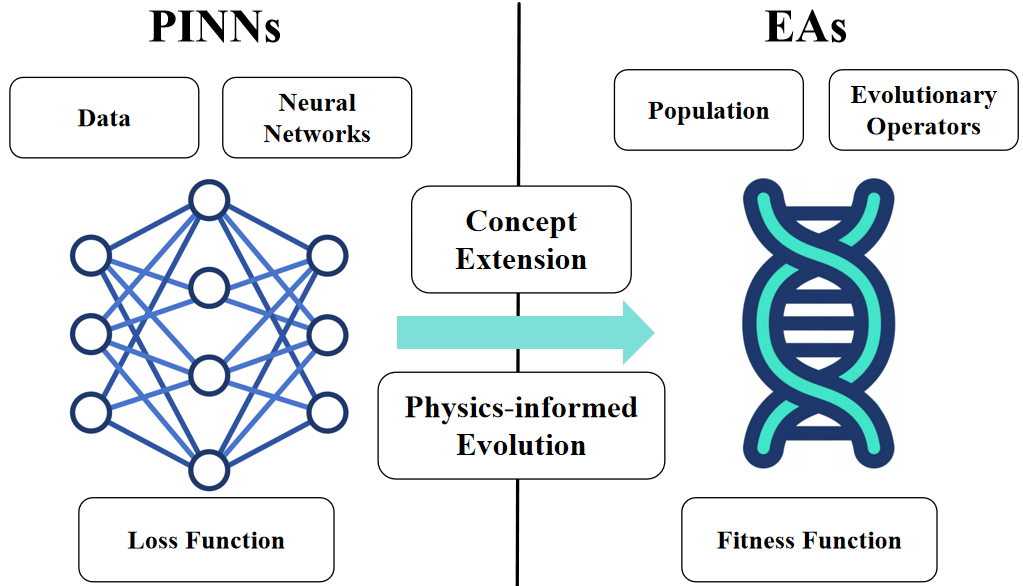}
	\caption{Extending physics-informed machine learning to physics-informed evolution. The figure illustrates the conceptual bridge between PINNs and PIE: physical information is embedded into the loss function in gradient-based learning, and analogously into the fitness function in population-based evolutionary search.}
	\label{fig:physics_to_evolution}
\end{figure}

This section reviews the two foundational areas underlying our work: physics-informed learning and evolutionary computation, with a focus on how prior knowledge can be integrated into optimization processes. Figure \ref{fig:physics_to_evolution} illustrates the conceptual extension from physics-informed machine learning to physics-informed evolution, highlighting the analogous role of physical information in guiding gradient-based learning and population-based evolutionary search.

\subsection{Physics-informed Neural Networks}

The core idea of PINNs is to embed the residual of a governing partial differential equation into the training objective of a neural network. Given a PDE of the form $\mathcal{F}[u] = 0$ defined over a spatial-temporal domain, a PINN parameterizes the solution $u$ as a neural network $u_\theta$ and minimizes a composite loss function:
\begin{equation}
	\mathcal{L}(\theta) = \mathcal{L}_{\text{data}}(\theta) + \lambda \mathcal{L}_{\text{PDE}}(\theta),
\end{equation}
where $\mathcal{L}_{\text{data}}(\theta) = \|u_\theta - u_{\text{obs}}\|^2$ measures the discrepancy between the network output and available observations, and $\mathcal{L}_{\text{PDE}}(\theta) = \|\mathcal{F}[u_\theta]\|^2$ penalizes violations of the governing equation at a set of collocation points, with $\lambda$ serving as a hyperparameter balancing the contribution of the two terms. The physical law acts as a form of structured regularization: it constrains the solution space to be physically consistent, enabling accurate solutions even when observational data is sparse or noisy.

Since Raissi et al. introduced the PINN framework in 2019, significant progress has been made in this field. Wang et al. systematically analyzed the training failure mechanisms of PINNs from the neural tangent kernel perspective\cite{wang2022and}, revealing that gradient flow pathologies are the root cause of training difficulties, and proposed mitigation strategies based on adaptive weighting and neural tangent kernel guidance. In the area of operator learning, Lu et al. proposed DeepONet\cite{lu2021learning}, which leverages the universal approximation theorem of operators to enable mapping from function spaces to function spaces, allowing the learning of relationships between different input functions. Li et al. introduced the Fourier Neural Operator, which parameterizes kernel integrals in Fourier space, enabling efficient solving of families of PDEs and significantly improving computational efficiency\cite{li2020fourier}. In recent years, the physics-informed paradigm has been integrated with neural operators, with Li et al. proposing the Physics-informed Neural Operator\cite{li2024physics}, which further incorporates physical constraints into the operator learning framework, ensuring that predictions not only learn the solution operator but also satisfy physical laws. Additionally, Kolmogorov-Arnold Networks have emerged as a promising neural architecture in scientific computing\cite{liu2024kan}. These works collectively establish the foundational role of physics-informed methods in the field of scientific machine learning.

\subsection{Evolutionary Algorithms}

EAs are population-based, derivative-free optimizers inspired by natural selection. Early representatives include Genetic Algorithms (GA)\cite{holland1992genetic}, which iteratively optimize populations through selection, crossover, and mutation operators. Single-objective EAs encompass Differential Evolution (DE)\cite{storn1997differential} and its adaptive variants such as Self-adaptive Differential Evolution (SaDE)\cite{qin2005self}, Adaptive Differential Evolution with Optional External Archive (JADE)\cite{zhang2009jade}, Success-History Based Parameter Adaptation for Differential Evolution (SHADE)\cite{tanabe2013success}, and Improving the Search Performance of SHADE Using Linear Population Size Reduction (LSHADE)\cite{tanabe2014improving}. Evolution Strategies include Covariance Matrix Adaptation Evolution Strategy (CMA-ES)\cite{hansen2003reducing}, Mixture Model-Based Evolution Strategy (MMES)\cite{he2020mmes}, and Evolution Strategy Based on Search Direction Adaptation (SDAES)\cite{he2019large}. Multi-objective EAs include Nondominated Sorting Genetic Algorithm II (NSGA-II)\cite{deb2002fast}, Reference-point-based Many Objective Evolutionary Algorithm Following NSGA-II (NSGA-III)\cite{deb2013evolutionary}, A Multiobjective Evolutionary Algorithm Based on Decomposition (MOEA/D)\cite{zhang2007moea}, Strength Pareto Evolutionary Algorithm 2 (SPEA2)\cite{zitzler2001spea2}, and Modified Pareto Envelop-based Selection Algorithm (PESA2)\cite{corne2000pareto,algorithm2013ipesa}.

The relationship between learning and evolution has been foundational in artificial intelligence. The Baldwin effect reveals how learning guides evolution by influencing gene frequency distributions\cite{weber2003evolution}. Memetic algorithms further embed local search methods such as hill climbing and gradient descent as ``memetic learning'' operators into the evolutionary framework\cite{ong2010memetic}.  Meanwhile, researchers have leveraged methods such as reinforcement learning, LLMs and graph neural networks to guide the design of evolutionary operators\cite{li2024bridging,hao2024large,ouyang2026learn}. These works demonstrate that learning and evolution are complementary paradigms that can mutually enhance each other. This connection naturally extends to the incorporation of prior knowledge: just as physics-informed learning embeds physical laws into loss functions, the same principle transfers to evolutionary computation, where physical information shapes the fitness landscape to guide gradient-free search. The Physics-informed Evolution framework establishes precisely this bridge between the two paradigms.

\section{Quantum Control Problems}
Quantum control aims to manipulate the dynamics of a quantum system by adjusting external control fields, enabling reliable quantum computation and information processing. The optimization objective is to find control functions that drive the system from an initial state $|\psi(0)\rangle$ to a target state $|\psi_{\text{target}}\rangle$ within time $T$, typically maximizing fidelity or minimizing state deviation\cite{dong2015sampling,li2021neural}.

The quantum state $|\psi(t)\rangle$ evolves according to the Schr\"odinger equation:
\begin{equation}
	\label{eq:schrodinger}
	i\hbar \frac{d}{dt}|\psi(t)\rangle = H(t)|\psi(t)\rangle
\end{equation}

The Hamiltonian $H(t)$ consists of contributions from control fields and inherent system dynamics. For a system with $M$ control fields and $F$ free or coupling fields, the Hamiltonian takes the linear form:
\begin{equation}
	\label{eq:linear_H}
	H(t) = \sum_{m=1}^{M} u_m(t)H_m + \sum_{f=1}^{F} H_f
\end{equation}
where $u_m(t)$ denotes the control functions to be optimized, and $H_m$ and $H_f$ denote the corresponding Hamiltonians. In many quantum systems, uncertain parameters $\{\theta_i\}$ affect the Hamiltonian through influence functions $f_i(t;\theta_i)$:
\begin{equation}
	\label{eq:uncertain_H}
	H(t) = \sum_{m=1}^{M} u_m(t)f_m(t;\theta_m)H_m + \sum_{f=1}^{F} f_{M+f}(t;\theta_{M+f})H_f
\end{equation}

\begin{figure}[t]
	\centering
	\includegraphics[width=0.7\textwidth]{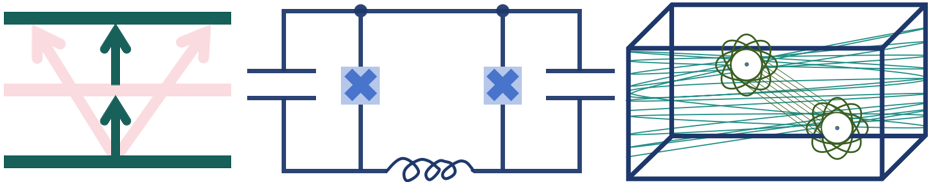}
	\caption{Illustration of the three quantum control problems: from left to right, the V-type three-level quantum system, superconducting quantum circuits, and two two-level atoms interacting with a quantized field.}
	\label{fig:qcp}
\end{figure}

The following three quantum control problems serve as benchmarks, representing different physical settings and increasing complexity. Figure \ref{fig:qcp} illustrates the schematic diagrams of these three problems.

\subsection{State Preparation in V-Type Three-Level Quantum Systems}
A V-type three-level system consists of two upper energy levels coupled to a common lower level, serving as a fundamental model for light-matter interactions and quantum interference. The system is initialized in the ground state and driven to a superposition target state:
\begin{equation}
	\label{eq:initial_state1}
	|\psi(0)\rangle = |g\rangle = (1, 0, 0), \qquad
	|\psi_{\text{target}}\rangle = \frac{1}{\sqrt{2}}(|e_1\rangle + |e_2\rangle) = \left(0, \frac{1}{\sqrt{2}}, \frac{1}{\sqrt{2}}\right)
\end{equation}

The Hamiltonian includes one free field and four control fields:
\begin{equation}
	\label{eq:hamiltonian_def}
	H(t) = f_0(t;\theta_0)H_0 + \sum_{m=1}^{4} u_m(t)f_m(t;\theta_m)H_m
\end{equation}

The Hamiltonian matrices are:
\begin{equation}
	\label{eq:matrices}
	\begin{aligned}
		& H_0 = \begin{pmatrix} 1.5 & 0 & 0 \\ 0 & 1 & 0 \\ 0 & 0 & 1 \end{pmatrix}, 
		H_1 = \begin{pmatrix} 0 & 1 & 0 \\ 1 & 0 & 0 \\ 0 & 0 & 0 \end{pmatrix}, 
		H_2 = \begin{pmatrix} 0 & -i & 0 \\ i & 0 & 0 \\ 0 & 0 & 0 \end{pmatrix}, \\
		& H_3 = \begin{pmatrix} 0 & 0 & 1 \\ 0 & 0 & 0 \\ 1 & 0 & 0 \end{pmatrix}, 
		H_4 = \begin{pmatrix} 0 & 0 & -i \\ 0 & 0 & 0 \\ i & 0 & 0 \end{pmatrix}
	\end{aligned}
\end{equation}

The influence functions are $f_0(t;\theta_0) = 1 - \epsilon \theta \cos t$  and $f_i(t;\theta_i) = 1$ for $i=1,\dots,4$, where $\epsilon$ is a system parameter.

\subsection{Superconducting Quantum Circuits}
Superconducting circuits based on Josephson junctions behave as artificial atoms and are promising platforms for scalable quantum computing. Qubits are controlled by external parameters such as gate voltage $V_g$ and magnetic flux $\Phi$. The Hamiltonian for a coupled two-qubit system with an LC oscillator, after normalization, takes the form:
\begin{equation}
	\label{eq:normalized_H}
	\begin{split}
		\frac{H}{\hbar} &= \theta_1 u_1(t)\sigma_{\zeta}(1)\otimes I(2) + \theta_2 u_2(t)\sigma_{\zeta}(1)\otimes \sigma_{\zeta}(2) - \theta_3 u_3(t)\sigma_x(1)\otimes I(2) \\
		&\quad - \theta_4 u_4(t)\sigma_x(1)\otimes \sigma_{\zeta}(2) - \theta_5 u_5(t)\sigma_y(1)\otimes \sigma_y(2)
	\end{split}
\end{equation}

Here $\theta_m$ denote uncertain coefficients, $u_i(t)$ denote control functions, and $\sigma$ denotes the Pauli matrices. The control objective is to drive the system from the ground state to a maximally entangled Bell state:
\begin{equation}
	\label{eq:initial_state3}
	|\psi(0)\rangle = |g_1 g_2\rangle = (1,0,0,0),
	|\psi_{\text{target}}\rangle = \tfrac{1}{\sqrt{2}}(|g_1 e_2\rangle + |e_1 g_2\rangle) = \bigl(0,\tfrac{1}{\sqrt{2}},\tfrac{1}{\sqrt{2}},0\bigr)
\end{equation}

\subsection{Two Two-Level Atoms Interacting with a Quantized Field}
This system models two atoms coupled to a cavity field, providing insights into entangled state dynamics essential for quantum communication. The total Hamiltonian comprises free evolution $H_0$, atom-atom and atom-field interactions $H_I$, and control terms $H_u$:
\begin{align}
\small
	\label{eq:H_0}
	H_0 &= \frac{1}{2}\sum_{i=1}^{2} \omega_{A_i}\sigma_z^{(i)} + \omega_r a^\dagger a, \\
	\label{eq:H_I}
	H_I &= \sum_{i \neq j} \Omega_{ij}\sigma_+^{(i)} \otimes \sigma_-^{(j)} + \sum_{j} v_j \left( a^\dagger \sigma_-^{(j)} + a \sigma_+^{(j)} \right), \\
	\label{eq:H_u}
	H_u &= \sum_{i=1}^{2} u_{\omega_{A_i}}(t)\sigma_z^{(i)} + u_{\omega_r}(t)a^\dagger a \notag \\
	&\quad + \sum_{i \neq j} u_{\Omega_{ij}}(t)\sigma_+^{(i)} \otimes \sigma_-^{(j)} + \sum_{j} u_{v_j}(t) \left( a^\dagger \sigma_-^{(j)} + a \sigma_+^{(j)} \right)
\end{align}
where $\omega_{A_i}$ and $\omega_r$ denote the atomic transition frequencies and the field frequency, respectively, $\Omega_{ij}$ denotes the dipole-dipole coupling, and $v_j$ denotes the atom-field coupling constant. The control task is to generate a maximally entangled state from the ground state:
\begin{equation}
	\label{eq:initial_state}
    \small
	|\psi(0)\rangle = |g_1 g_2\rangle = (1,0,0,0),
	|\psi_{\text{target}}\rangle = \tfrac{1}{\sqrt{2}}(|g_1 e_2\rangle + |e_1 g_2\rangle) = \bigl(0,\tfrac{1}{\sqrt{2}},\tfrac{1}{\sqrt{2}},0\bigr)
\end{equation}

\section{Physics-informed Quantum Control}
This section applies the Physics-informed Evolution framework to quantum control. We detail encoding control fields into evolutionary search spaces and constructing fitness functions that balance state fidelity with energy constraints, covering both single- and multi-objective formulations.

\subsection{Encoding of Search Space}

The control functions $u$ are discretized via uniform grid sampling. For a total control time $T$, we take $N$ sample points $\{t_0, t_1, \dots, t_N\}$ with $t_i = iT/N$, and the values at the first $N$ points form the search space. For $M$ control functions, the search space has dimension $M \times N$, as defined in equation \eqref{eq:Pos}.

\begin{equation}
\small
	\label{eq:Pos}
	\begin{aligned}
		Pos(i) = \{ & u_1(t_0), \dots, u_1(t_{N-1}), u_2(t_0), \dots, u_2(t_{N-1}), \\
		& \dots, u_M(t_0), \dots, u_M(t_{N-1}) \}
	\end{aligned}
\end{equation}

Given the control functions and uncertain parameters, the Schr\"odinger equation \eqref{eq:ihbar} is solved numerically using the finite difference method:
{\footnotesize
\begin{equation}
	\label{eq:ihbar}
	i\hbar \frac{d}{dt}|\psi(t)\rangle = H(t)|\psi(t)\rangle = \left( \sum_{i=1}^M u_i(t)f_i(t;\theta_i)H_i(t)+\sum_{j=M+1}^{M+F}f_j(t;\theta_j)H_j(t)\right) |\psi(t)\rangle 
\end{equation}
}
The states at sample points follow the recursion in equation \eqref{eq:Psi}, enabling computation of the full trajectory $\{\Psi(t_i)\}_{i=0}^N$:
\begin{equation}
	\label{eq:Psi}
	\Psi(t_{i+1}) = \Psi(t_i) - \frac{i}{\hbar}H(t_i)\Psi(t_i)\Delta t 
\end{equation}
We use spline interpolation to balance accuracy and dimensionality. By interpolating $N$ search samples into $\alpha(N-1)$ points, we ensure simulation precision without increasing optimization variables.

\begin{figure}[t]
	\centering
	\includegraphics[width=0.7\textwidth]{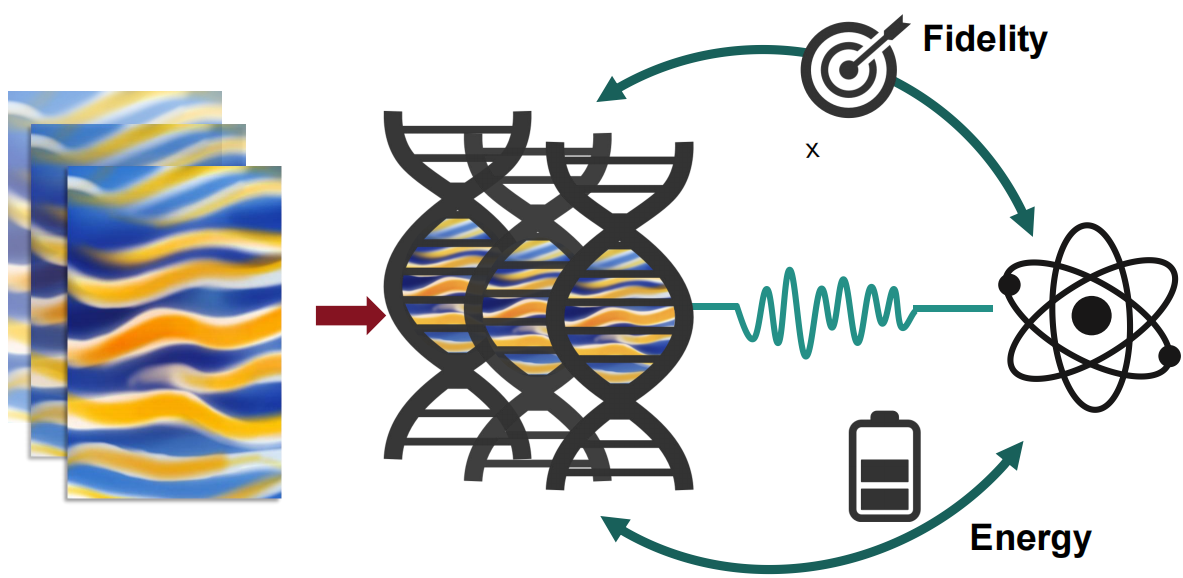}
	\caption{The framework of PIE. Control fields are encoded via grid sampling with spline interpolation. Physical information is embedded into the fitness function by incorporating state deviation and energy consumption, guiding evolutionary search toward physically consistent solutions.}
	\label{fig:pie_framework}
\end{figure}
\subsection{Physics-informed Evolution (PIE)}

The primary optimization objective is fidelity, defined as the overlap between the final state $\Psi(T)$ and the target state $\Psi_{\text{target}}$. For pure states:
\begin{equation}
	\label{eq:F_fidelity}
	F(u) = |\langle\Psi(T)|\Psi_{\text{target}}\rangle|^2 
\end{equation}
For mixed states, fidelity is computed via the trace distance between density matrices $\rho(T) = |\Psi(T)\rangle\langle\Psi(T)|$ and $\rho_{\text{target}} = |\Psi_{\text{target}}\rangle\langle\Psi_{\text{target}}|$:
\begin{equation}
	\label{eq:F_fidelity_mixed}
	F(u) = \text{Tr} \left( \sqrt{\sqrt{\rho(T)} \rho_{\text{target}} \sqrt{\rho(T)}} \right)
\end{equation}

Since fidelity maximization can lead to optimization difficulties (optimal solutions cluster in distant regions), we instead use state deviation as the objective. For pure states:
\begin{equation}
	\label{eq:D_pure}
	D(u) = \| \Psi(T) - \Psi_{\text{target}} \| 
\end{equation}
For mixed states:
\begin{equation}
	\label{eq:D_mixed}
	D(u) = \| \rho(T) - \rho_{\text{target}} \| 
\end{equation}
Minimizing $D(u)$ is equivalent to maximizing fidelity, referred to as negative fidelity optimization.

Energy consumption, reflecting control cost, is defined as:
\begin{equation}
	\label{eq:Energy}
	E(u) = \sum_{i=1}^{M} \int_{0}^{T} |u_i(t)| \cdot \| H_i(t) \| dt = \sum_{i=1}^{M} \sum_{j=1}^{N} |u_i(t_j)| \cdot \| H_i(t_j) \| \cdot \Delta t
\end{equation}

Drawing inspiration from PINNs, in which the loss function combines data fidelity with physics-informed regularization, we construct the PIE fitness function in an analogous manner. The state deviation $D(u)$ serves as the \textit{data loss}, quantifying task performance, while the energy consumption $E(u)$ serves as the \textit{physics-informed regularization}, reflecting physical priors on control cost. The single-objective formulation combines the two terms through a weighted sum:
\begin{equation}
	\label{eq:fitness}
	J(u) = D(u) + \lambda \cdot E(u)
\end{equation}
where $\lambda$ balances fidelity and energy efficiency.

Alternatively, a multi-objective formulation treats $D(u)$ and $E(u)$ as separate objectives to be minimized simultaneously, eliminating the need for $\lambda$ tuning and revealing the Pareto-optimal trade-offs:
\begin{equation}
	\label{eq:mop}
	\begin{aligned}
		& \underset{u}{\text{minimize}} & & \mathbf{J}(u) = \big( D(u), E(u) \big)
	\end{aligned}
\end{equation}
The resulting Pareto front provides a range of control strategies, from high-fidelity to energy-efficient solutions, without requiring a priori preference specification. Fig. \ref{fig:pie_framework} illustrates the overall PIE framework.
\section{Experimental Results and Analysis}
\subsection{Experimental Setup}
To comprehensively evaluate the proposed method, widely used evolutionary optimizers are selected as baselines. The single-objective algorithms include DE\cite{storn1997differential}, CMA-ES\cite{hansen2003reducing}, MMES\cite{he2020mmes}, SDAES\cite{he2019large}, SaDE\cite{qin2005self}, JADE\cite{zhang2009jade}, CJADE, SHADE\cite{tanabe2013success}, LSHADE\cite{tanabe2014improving}, and EA4eig. The multiobjective algorithms include NSGA-II\cite{deb2002fast}, NSGA-III\cite{deb2013evolutionary}, MOEA/D\cite{zhang2007moea}, SPEA2\cite{zitzler2001spea2}, and PESA2\cite{corne2000pareto,algorithm2013ipesa}. These baselines cover differential evolution variants, evolution strategies, and representative multiobjective frameworks.

For a fair comparison, all algorithms use a population size $N=50$, $T=500$ iterations, and 30 independent runs. Parameters follow Table~\ref{tab:params} and original implementations to avoid adjustment bias. Experiments are conducted on an Intel Core i5-12400 CPU (2.50 GHz) with 16 GB RAM, using MATLAB R2023a.

\begin{table}[t]
	\centering
	\caption{Parameter configurations for competing algorithms}
	\label{tab:params}
	\renewcommand{\arraystretch}{1.0} 
    \setlength{\aboverulesep}{2pt} 
    \setlength{\belowrulesep}{2pt} 
	\begin{tabular}{lp{10cm}}
		\toprule
		\textbf{Algorithms} & \textbf{Specific parameters} \\
		\midrule
		DE & $F=0.5, CR=0.5$ \\
		CMA-ES & $\lambda = 4 + \lfloor 3 \ln n \rfloor, \mu = \lfloor \lambda/2 \rfloor, \omega_i = \frac{\ln(\mu+0.5)-\ln(i)}{\mu \ln(\mu+0.5)-\sum_{j=1}^{\mu} \ln(j)}, \quad \mu_{eff} = \frac{1}{\sum_{i=1}^{\mu} \omega_i^2}, c_c = 0.4/\sqrt{n}$ \\
		MMES & $\lambda = 4 + \lfloor 3 \ln n \rfloor, \mu = \lfloor \lambda/2 \rfloor, m = 2 \lceil \sqrt{n} \rceil, c_a = 4/n, c_c = 0.4/\sqrt{n}, T = \lceil 1/c_c \rceil, \gamma = 1 - (1 - c_a)^m, \omega_i = \frac{\ln(\mu+0.5)-\ln(i)}{\mu \ln(\mu+0.5)-\sum_{j=1}^{\mu} \ln(j)}, \quad \mu_{eff} = \frac{1}{\sum_{i=1}^{\mu} \omega_i^2}, c_\sigma = 0.3, d_\sigma = 1, \alpha_z = 0.05$ \\
		
		SDAES & $\lambda = 4 + \lfloor 3 \ln n \rfloor, \mu = \lfloor \lambda/2 \rfloor, m=10, c_{cov} = 0.4/\sqrt{n}, c_c = 0.25/\sqrt{n}, \omega_i = \frac{\ln(\mu+1)-\ln(i)}{\mu \ln(\mu+1)-\sum_{j=1}^{\mu} \ln(j)}, \quad \mu_{eff} = \frac{1}{\sum_{i=1}^{\mu} \omega_i^2}, c_s = 0.3, d_\sigma = 1, p^* = 0.05$ \\
		SaDE & $LP = 50, CR = \text{adaptive}$ \\
		JADE & $p=[0.05, 0.2], c=[0.05, 0.2], \mu_{CR}=0.5, \mu_{CF}=0.5$ \\
		CJADE & $Radius=0.001, c=0.1, p=0.05, CRm=0.5, Fm=0.5, A_{factor}=1$ \\
		SHADE & $NP=50, H=[30, 100]$ \\
		LSHADE & $NP_{init}=50, NP_{min}=4, p=0.11, |A|=2.6, H=6$ \\
		EA4eig & $NP_{init}=50, NP_{min}=10, p_{best\_rate}=0.11, |A|=2.6, H=4$ \\
		
		NSGA-II & $NP=50, p_c=0.9, p_m=1/D, \eta_c=20, \eta_m=20$ \\
		NSGA-III & $NP \approx H, p_c=0.9, p_m=1/D, \eta_c=30, \eta_m=20$ \\
		MOEA/D & $NP=50, T=20, \delta=0.9, n_r=2$ \\
		SPEA2 & $NP=50, \text{Archive}=50, p_c=0.9, p_m=1/D$ \\
		PESA2 & $NP=50, \text{Archive}=50, \text{Grid}=32, p_c=0.9, p_m=1/D$ \\
		\bottomrule
	\end{tabular}
\end{table}
\subsection{Experimental Results of Single-Objective Evolutionary Algorithms within the PIE Framework}

The single-objective PIE framework is evaluated using ten evolutionary algorithms on three quantum control scenarios: state preparation in V-type three-level quantum systems (S1), superconducting quantum circuits (S2), and two two-level atoms interacting with a quantized field (S3), under five regularization settings ($\lambda = 0, 0.1, 0.01, 0.001, 0.0001$). Table~\ref{tab:combined_results} presents the mean and standard deviation of Fidelity, Deviation, Energy, and RunTime over 30 independent runs for the case $\lambda = 0.1$. Results for other $\lambda$ values are provided in the supplementary materials, which are available at https://sandbox.zenodo.org/records/474222. The fitness function is defined as $J(u) = D(u) + \lambda \cdot E(u)$, where $\lambda$ controls the trade-off between state accuracy and energy consumption. Fig. \ref{fig:evolution} presents the quantum control results of the single-objective PIE framework across three problems, showing the evolution of the control field $u$ and the quantum state $\psi$ during the optimization process.

For $\lambda = 0.1$, where both state deviation and energy consumption are considered, distinct algorithm performances are observed across the three scenarios. In S1, DE achieves the highest fidelity of $9.20 \times 10^{-1}$ and the lowest deviation of $2.06 \times 10^{-2}$, outperforming other algorithms in state accuracy. MMES delivers the best performance in S2, attaining a fidelity of $9.38 \times 10^{-1}$ with a deviation of $1.92 \times 10^{-2}$. In S3, SDAES yields the highest fidelity of $8.63 \times 10^{-1}$ and the lowest deviation of $8.15 \times 10^{-2}$. Regarding energy consumption, SHADE achieves the most energy-efficient solutions in S1 ($2.01 \times 10^{0}$) and S3 ($3.64 \times 10^{0}$), while EA4eig provides the lowest energy consumption in S2 ($2.11 \times 10^{0}$). For runtime efficiency, EA4eig is the fastest in S1 with $5.92 \times 10^{1}$ seconds, while SDAES exhibits the shortest runtime in S2 ($4.60 \times 10^{1}$ seconds) and S3 ($6.65 \times 10^{1}$ seconds). These results demonstrate that when the energy regularization term is introduced at $\lambda = 0.1$, different algorithms excel in different scenarios, highlighting the inherent trade-off between state accuracy and energy efficiency. The performance variations across S1, S2, and S3 reflect the distinct physical characteristics and control complexities of each quantum system, underscoring the importance of algorithm selection tailored to specific problem settings.

\subsection{Experimental Results of Multi-Objective Evolutionary Algorithms within the PIE Framework}
\begin{figure}[!t]
	\centering
	\includegraphics[width=0.9\textwidth]{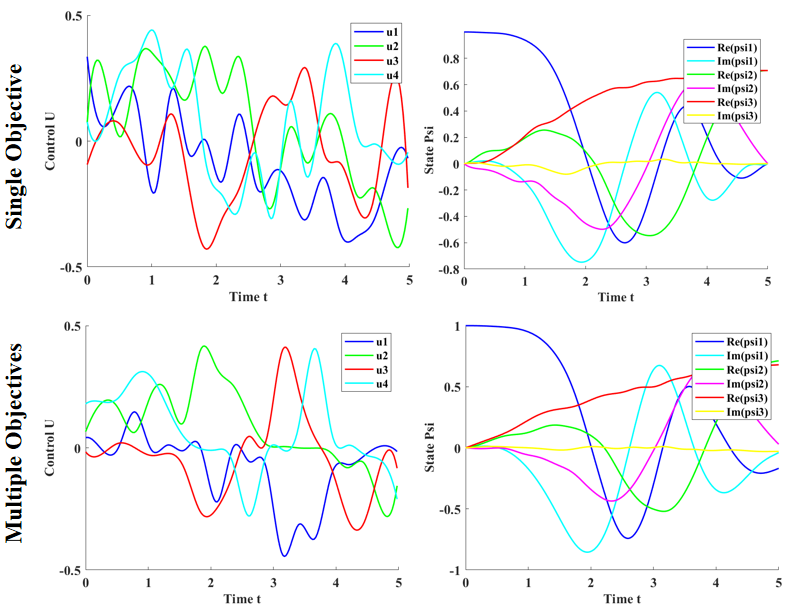}
	\caption{Evolution of control inputs $u$ and quantum state $\psi$ for three quantum control problems in the PIE framework.}
	\label{fig:evolution}
\end{figure}

\begin{table}[!t]
    \centering
    \caption{Comparison of single-objective and multi-objective evolutionary algorithms within the PIE framework.}
    \label{tab:combined_results}

    \tiny
    \setlength{\tabcolsep}{2pt}
    \renewcommand{\arraystretch}{0.9}
    \setlength{\aboverulesep}{1pt} 
    \setlength{\belowrulesep}{1pt} 
    \resizebox{\textwidth}{!}{%
    \begin{tabular}{c l l c c c}
        \toprule
        \textbf{Type} & \textbf{Algorithm} & \textbf{Metrics} & \textbf{S1} & \textbf{S2} & \textbf{S3} \\
        \midrule
        \multirow{40}{*}{\makecell{Single-\\Objective}} & \multirow{4}{*}{DE} & Fidelity & \textbf{9.20E-01$\pm$1.69E-02} & 7.72E-01$\pm$1.57E-02 & 7.87E-01$\pm$7.59E-02 \\
         & & Deviation & \textbf{2.06E-02$\pm$4.43E-03} & 7.92E-02$\pm$5.15E-03 & 1.32E-01$\pm$5.22E-02 \\
         & & Energy & 2.23E+00$\pm$4.41E-02 & 3.02E+00$\pm$7.40E-02 & 2.39E+01$\pm$8.64E-01 \\
         & & RunTime & 1.15E+03$\pm$5.46E+03 & 1.79E+02$\pm$4.59E+01 & 2.07E+02$\pm$7.16E+00 \\
        \cmidrule{2-6}
        & \multirow{4}{*}{CMA ES} & Fidelity & 9.16E-01$\pm$4.45E-03 & 9.15E-01$\pm$2.14E-03 & 8.16E-01$\pm$1.69E-02 \\
         & & Deviation & 2.17E-02$\pm$1.09E-03 & 3.23E-02$\pm$7.79E-04 & 9.23E-02$\pm$8.57E-03 \\
         & & Energy & 2.05E+00$\pm$9.80E-03 & 3.17E+00$\pm$8.15E-03 & 3.65E+00$\pm$2.98E-01 \\
         & & RunTime & 1.76E+02$\pm$5.29E+01 & 2.62E+02$\pm$2.32E+01 & 2.56E+02$\pm$1.51E+02 \\
        \cmidrule{2-6}
        & \multirow{4}{*}{MMES} & Fidelity & 7.92E-01$\pm$2.54E-01 & \textbf{9.38E-01$\pm$3.25E-02} & 8.16E-01$\pm$1.51E-01 \\
         & & Deviation & 6.84E-02$\pm$8.58E-02 & \textbf{1.92E-02$\pm$8.97E-03} & 9.98E-02$\pm$8.07E-02 \\
         & & Energy & 2.57E+01$\pm$1.18E+01 & 1.93E+01$\pm$8.14E+00 & 4.97E+01$\pm$1.41E+01 \\
         & & RunTime & 7.15E+01$\pm$2.09E+00 & 4.75E+01$\pm$1.21E+00 & 6.92E+01$\pm$1.64E+00 \\
        \cmidrule{2-6}
        & \multirow{4}{*}{SDAES} & Fidelity & 9.07E-01$\pm$1.37E-01 & 8.53E-01$\pm$1.91E-01 & \textbf{8.63E-01$\pm$8.96E-02} \\
         & & Deviation & 2.93E-02$\pm$4.05E-02 & 5.16E-02$\pm$8.96E-02 & \textbf{8.15E-02$\pm$5.13E-02} \\
         & & Energy & 1.90E+01$\pm$6.12E+00 & 1.13E+01$\pm$3.98E+00 & 3.37E+01$\pm$8.69E+00 \\
         & & RunTime & 7.09E+01$\pm$2.25E+00 & \textbf{4.60E+01$\pm$4.81E-01} & \textbf{6.65E+01$\pm$1.75E-01} \\
        \cmidrule{2-6}
        & \multirow{4}{*}{SaDE} & Fidelity & 9.14E-01$\pm$4.26E-03 & 6.59E-01$\pm$2.63E-01 & 7.70E-01$\pm$3.77E-02 \\
         & & Deviation & 2.21E-02$\pm$1.14E-03 & 1.39E-01$\pm$1.44E-01 & 1.25E-01$\pm$2.24E-02 \\
         & & Energy & 2.06E+00$\pm$1.35E-02 & 2.52E+00$\pm$1.00E+00 & 1.09E+01$\pm$2.65E+00 \\
         & & RunTime & 2.15E+02$\pm$2.06E+00 & 1.42E+02$\pm$5.22E-01 & 1.95E+02$\pm$3.51E-01 \\
        \cmidrule{2-6}
        & \multirow{4}{*}{JADE} & Fidelity & 9.16E-01$\pm$4.05E-03 & 7.16E-01$\pm$1.49E-01 & 7.62E-01$\pm$3.05E-02 \\
         & & Deviation & 2.17E-02$\pm$9.72E-04 & 1.04E-01$\pm$7.74E-02 & 1.26E-01$\pm$1.86E-02 \\
         & & Energy & 2.02E+00$\pm$9.60E-03 & 2.78E+00$\pm$5.32E-01 & 5.11E+00$\pm$3.03E+00 \\
         & & RunTime & 2.59E+02$\pm$4.93E+01 & 2.85E+02$\pm$7.02E-01 & 3.90E+02$\pm$5.01E+00 \\
        \cmidrule{2-6}
        & \multirow{4}{*}{CJADE} & Fidelity & 9.16E-01$\pm$3.23E-03 & 7.04E-01$\pm$1.66E-01 & 7.57E-01$\pm$2.86E-02 \\
         & & Deviation & 2.19E-02$\pm$9.03E-04 & 1.06E-01$\pm$8.01E-02 & 1.27E-01$\pm$1.70E-02 \\
         & & Energy & 2.02E+00$\pm$1.32E-02 & 2.75E+00$\pm$5.35E-01 & 4.06E+00$\pm$2.40E+00 \\
         & & RunTime & 1.74E+02$\pm$4.22E+01 & 1.53E+02$\pm$3.18E+00 & 1.98E+02$\pm$3.37E-01 \\
        \cmidrule{2-6}
        & \multirow{4}{*}{SHADE} & Fidelity & 9.17E-01$\pm$3.10E-03 & 7.06E-01$\pm$1.92E-01 & 7.60E-01$\pm$2.98E-02 \\
         & & Deviation & 2.21E-02$\pm$9.19E-04 & 1.12E-01$\pm$1.06E-01 & 1.24E-01$\pm$1.73E-02 \\
         & & Energy & \textbf{2.01E+00$\pm$1.39E-02} & 2.69E+00$\pm$7.31E-01 & \textbf{3.64E+00$\pm$1.88E+00} \\
         & & RunTime & 1.24E+02$\pm$2.54E+00 & 1.49E+02$\pm$2.76E-01 & 1.97E+02$\pm$3.31E-01 \\
        \cmidrule{2-6}
        & \multirow{4}{*}{LSHADE} & Fidelity & 9.17E-01$\pm$2.71E-03 & 7.57E-01$\pm$6.72E-03 & 7.67E-01$\pm$2.68E-02 \\
         & & Deviation & 2.15E-02$\pm$8.18E-04 & 8.40E-02$\pm$1.86E-03 & 1.23E-01$\pm$1.50E-02 \\
         & & Energy & 2.02E+00$\pm$1.14E-02 & 2.89E+00$\pm$1.87E-02 & 3.86E+00$\pm$1.27E+00 \\
         & & RunTime & 1.21E+02$\pm$3.32E-01 & 1.48E+02$\pm$2.62E-01 & 1.95E+02$\pm$3.16E-01 \\
        \cmidrule{2-6}
        & \multirow{4}{*}{EA4eig} & Fidelity & 8.80E-01$\pm$1.65E-01 & 3.67E-01$\pm$2.40E-01 & 5.66E-01$\pm$1.98E-01 \\
         & & Deviation & 4.05E-02$\pm$9.53E-02 & 2.86E-01$\pm$1.40E-01 & 5.14E-01$\pm$1.99E-01 \\
         & & Energy & 2.44E+00$\pm$2.27E-01 & \textbf{2.11E+00$\pm$7.37E-01} & 4.97E+00$\pm$2.32E+01 \\
         & & RunTime & \textbf{5.92E+01$\pm$2.28E+00} & 6.53E+01$\pm$1.28E+01 & 7.94E+01$\pm$1.73E+01 \\
        \midrule
        \multirow{20}{*}{\makecell{Multi-\\Objectives}} & \multirow{4}{*}{NSGA-II} & Fidelity & 8.58E-01$\pm$8.29E-02 & \textbf{1.00E+00$\pm$1.49E-05} & 9.40E-01$\pm$4.32E-02 \\
         & & Deviation & 3.74E-02$\pm$2.23E-02 & \textbf{5.47E-06$\pm$3.66E-06} & 3.00E-02$\pm$2.16E-02 \\
         & & Energy & \textbf{2.10E+00$\pm$2.34E-01} & 3.32E+01$\pm$2.68E+00 & 7.39E+01$\pm$1.25E+01 \\
         & & RunTime & 1.60E+02$\pm$4.08E+00 & 1.74E+02$\pm$1.26E+00 & 2.43E+02$\pm$4.56E+00 \\
        \cmidrule{2-6}
        & \multirow{4}{*}{NSGA-III} & Fidelity & 9.32E-01$\pm$5.00E-02 & 9.99E-01$\pm$1.22E-03 & 9.39E-01$\pm$3.33E-02 \\
         & & Deviation & 1.76E-02$\pm$1.30E-02 & 1.44E-04$\pm$3.04E-04 & 3.06E-02$\pm$1.66E-02 \\
         & & Energy & 2.25E+00$\pm$1.71E-01 & \textbf{3.23E+01$\pm$1.61E+00} & 7.04E+01$\pm$1.18E+01 \\
         & & RunTime & 1.48E+02$\pm$6.52E-01 & 1.64E+02$\pm$8.70E-01 & 2.32E+02$\pm$8.35E-01 \\
        \cmidrule{2-6}
        & \multirow{4}{*}{MOEA/D} & Fidelity & 9.27E-01$\pm$3.07E-02 & 9.53E-01$\pm$3.76E-02 & \textbf{9.64E-01$\pm$2.15E-02} \\
         & & Deviation & 2.10E-02$\pm$9.59E-03 & 2.14E-02$\pm$2.90E-02 & \textbf{1.79E-02$\pm$1.09E-02} \\
         & & Energy & 3.00E+00$\pm$2.26E-01 & 4.52E+01$\pm$7.87E+00 & 9.21E+01$\pm$3.78E+00 \\
         & & RunTime & 1.50E+02$\pm$1.38E+01 & 1.42E+02$\pm$4.12E+00 & 2.15E+02$\pm$1.98E+00 \\
        \cmidrule{2-6}
        & \multirow{4}{*}{SPEA2} & Fidelity & 9.52E-01$\pm$3.33E-02 & 8.47E-01$\pm$1.61E-01 & 9.27E-01$\pm$2.36E-02 \\
         & & Deviation & 1.25E-02$\pm$8.62E-03 & 4.35E-02$\pm$4.57E-02 & 3.86E-02$\pm$1.26E-02 \\
         & & Energy & 2.76E+00$\pm$3.29E-01 & 3.29E+01$\pm$7.15E+00 & \textbf{6.43E+01$\pm$7.36E+00} \\
         & & RunTime & 1.44E+02$\pm$1.35E+00 & 1.45E+02$\pm$4.73E-01 & 2.10E+02$\pm$7.31E-01 \\
        \cmidrule{2-6}
        & \multirow{4}{*}{PESA2} & Fidelity & \textbf{9.89E-01$\pm$4.23E-03} & 7.95E-01$\pm$9.41E-02 & 9.29E-01$\pm$3.51E-02 \\
         & & Deviation & \textbf{2.82E-03$\pm$1.12E-03} & 6.18E-02$\pm$2.78E-02 & 4.15E-02$\pm$1.88E-02 \\
         & & Energy & 3.10E+00$\pm$1.30E-01 & 3.49E+01$\pm$5.12E+00 & 7.44E+01$\pm$5.92E+00 \\
         & & RunTime & \textbf{1.30E+02$\pm$2.32E+00} & \textbf{1.38E+02$\pm$7.74E-01} & \textbf{2.01E+02$\pm$2.53E+00} \\
        \bottomrule
    \end{tabular}%
    }
\end{table}

The multi-objective PIE framework is evaluated by comparing five representative multi-objective evolutionary algorithms, namely NSGA-II, NSGA-III, MOEA/D, SPEA2, and PESA2, across the three quantum control scenarios S1, S2, and S3.For each algorithm, 30 independent runs are conducted. From the Pareto front obtained in each run, the solution with the smallest Deviation is selected as the preferred solution. This criterion aligns with practical quantum control applications, where state accuracy is typically prioritized over energy efficiency. Table \ref{tab:combined_results} summarizes the mean and standard deviation of Fidelity, Deviation, Energy, and RunTime for the preferred solutions over all runs.
Fig. \ref{fig:evolution} shows the quantum control results from the multi-objective PIE framework, illustrating the evolution of control field $u$ and quantum state $\psi$ for the preferred Pareto solutions. Compared to the single-objective approach, the multi-objective control fields consume less energy and yield smoother, more accurate state evolution.

In S1, PESA2 achieves the highest fidelity of $9.89 \times 10^{-1}$ and the lowest deviation of $2.82 \times 10^{-3}$, significantly outperforming other algorithms. NSGA-II exhibits the best energy efficiency ($2.10 \times 10^{0}$) but with lower fidelity ($8.58 \times 10^{-1}$). SPEA2 offers a balanced trade-off with fidelity $9.52 \times 10^{-1}$ and energy $2.76 \times 10^{0}$. PESA2 is also the fastest, with a runtime of $1.30 \times 10^{2}$ seconds.In S2, NSGA-II delivers near-perfect fidelity of $1.00 \times 10^{0}$ with deviation $5.47 \times 10^{-6}$, while NSGA-III achieves the lowest energy ($3.23 \times 10^{1}$) with fidelity $9.99 \times 10^{-1}$. PESA2 is the fastest ($1.38 \times 10^{2}$ seconds) but yields lower fidelity ($7.95 \times 10^{-1}$).In S3, MOEA/D achieves the best state accuracy with fidelity $9.64 \times 10^{-1}$ and deviation $1.79 \times 10^{-2}$, while SPEA2 attains the lowest energy ($6.43 \times 10^{1}$). PESA2 is again the fastest ($2.01 \times 10^{2}$ seconds).

Compared to the single-objective weighted-sum formulation at $\lambda=0.1$ (Table~\ref{tab:combined_results}), the multi-objective approach offers distinct advantages. In the single-objective setting, the regularization parameter $\lambda$ requires careful tuning to balance state accuracy and energy efficiency, and the chosen $\lambda=0.1$ yields different optimal algorithms across scenarios: DE in S1 (fidelity $9.20 \times 10^{-1}$), MMES in S2 (fidelity $9.38 \times 10^{-1}$), and SDAES in S3 (fidelity $8.63 \times 10^{-1}$). In contrast, the multi-objective framework eliminates $\lambda$ tuning by directly exploring the Pareto front, allowing decision-makers to select solutions that best suit their preferences. Moreover, the multi-objective results achieve higher state accuracy in S1 ($9.89 \times 10^{-1}$ vs. $9.20 \times 10^{-1}$) and S3 ($9.64 \times 10^{-1}$ vs. $8.63 \times 10^{-1}$) compared to the best single-objective solutions, demonstrating that the multi-objective formulation can discover superior trade-offs by avoiding the sensitivity to hyperparameter selection inherent in weighted-sum approaches.

Overall, no single algorithm dominates across all scenarios: PESA2 performs best in S1, NSGA-II in S2, and MOEA/D in S3 when state accuracy is prioritized. The Deviation–Energy trade-off is consistently evident, and the multi-objective PIE framework successfully produces diverse Pareto-optimal solutions that offer greater flexibility than the fixed-weight single-objective formulation.

\section{Conclusions and Future Work}

This paper introduces PIE, a framework that embeds physical information from governing equations into the fitness function of EAs. Drawing a direct analogy with PINNs, we demonstrate that physical insights guiding gradient-based learning can similarly shape evolutionary search. By incorporating energy consumption as a physics-informed regularization term in the single-objective formulation or as a competing objective in the multi-objective formulation, PIE guides evolutionary algorithms toward solutions that are both high-performing and physically consistent. We validate PIE on three quantum control problems (S1, S2, S3) governed by the Schr\"odinger equation. Extensive experiments compare ten single-objective and five multi-objective evolutionary baselines. In the single-objective setting with $\lambda=0.1$, DE, MMES, and SDAES achieve the highest fidelity in S1, S2, and S3, respectively. In the multi-objective setting, PESA2, NSGA-II, and MOEA/D achieve the best state accuracy in S1, S2, and S3, respectively. Compared to the weighted-sum approach, the multi-objective formulation eliminates manual $\lambda$ tuning and discovers superior trade-offs. Beyond empirical validation, this work establishes a conceptual bridge between physics-informed machine learning and evolutionary computation, demonstrating that embedding physical information into optimization objectives extends naturally to population-based search. Future directions include developing differentiable evolutionary frameworks, incorporating physical principles into evolutionary operators, and extending PIE to other problems governed by physical laws, such as flow control.

\begin{credits}
\subsubsection{\ackname} This work was supported by the National Natural Science Foundation of China (U25A20450, 62571374).

\subsubsection{\discintname} The authors declare that they have no conflicts of interest.
 
\end{credits}
%
%
%
\bibliographystyle{splncs04}
\bibliography{ref}

\begin{thebibliography}{10}
\providecommand{\url}[1]{\texttt{#1}}
\providecommand{\urlprefix}{URL }
\providecommand{\doi}[1]{https://doi.org/#1}

\bibitem{algorithm2013ipesa}
Algorithm~II, S.: Ipesa-ii: Improved pareto envelope-based. In: Evolutionary
  Multi-Criterion Optimization: 7th International Conference, EMO 2013,
  Sheffield, UK, March 19-22, 2013. Proceedings. p.~143. Springer (2013)

\bibitem{bottou2018optimization}
Bottou, L., Curtis, F.E., Nocedal, J.: Optimization methods for large-scale
  machine learning. SIAM review  \textbf{60}(2),  223--311 (2018)

\bibitem{corne2000pareto}
Corne, D.W., Knowles, J.D., Oates, M.J.: The pareto envelope-based selection
  algorithm for multiobjective optimization. In: International conference on
  parallel problem solving from nature. pp. 839--848. Springer (2000)

\bibitem{deb2013evolutionary}
Deb, K., Jain, H.: An evolutionary many-objective optimization algorithm using
  reference-point-based nondominated sorting approach, part i: solving problems
  with box constraints. IEEE transactions on evolutionary computation
  \textbf{18}(4),  577--601 (2013)

\bibitem{deb2002fast}
Deb, K., Pratap, A., Agarwal, S., Meyarivan, T.: A fast and elitist
  multiobjective genetic algorithm: Nsga-ii. IEEE transactions on evolutionary
  computation  \textbf{6}(2),  182--197 (2002)

\bibitem{dong2015sampling}
Dong, D., Mabrok, M.A., Petersen, I.R., Qi, B., Chen, C., Rabitz, H.:
  Sampling-based learning control for quantum systems with uncertainties. IEEE
  Transactions on Control Systems Technology  \textbf{23}(6),  2155--2166
  (2015)

\bibitem{hansen2003reducing}
Hansen, N., M{\"u}ller, S.D., Koumoutsakos, P.: Reducing the time complexity of
  the derandomized evolution strategy with covariance matrix adaptation
  (cma-es). Evolutionary computation  \textbf{11}(1),  1--18 (2003)

\bibitem{hao2024large}
Hao, H., Zhang, X., Zhou, A.: Large language models as surrogate models in
  evolutionary algorithms: A preliminary study. Swarm and Evolutionary
  Computation  \textbf{91},  101741 (2024)

\bibitem{he2020mmes}
He, X., Zheng, Z., Zhou, Y.: Mmes: Mixture model-based evolution strategy for
  large-scale optimization. IEEE Transactions on Evolutionary Computation
  \textbf{25}(2),  320--333 (2020)

\bibitem{he2019large}
He, X., Zhou, Y., Chen, Z., Zhang, J., Chen, W.N.: Large-scale evolution
  strategy based on search direction adaptation. IEEE Transactions on
  Cybernetics  \textbf{51}(3),  1651--1665 (2019)

\bibitem{holland1992genetic}
Holland, J.H.: Genetic algorithms. Scientific american  \textbf{267}(1),
  66--73 (1992)

\bibitem{karniadakis2021physics}
Karniadakis, G.E., Kevrekidis, I.G., Lu, L., Perdikaris, P., Wang, S., Yang,
  L.: Physics-informed machine learning. Nature Reviews Physics  \textbf{3}(6),
   422--440 (2021)

\bibitem{kingma2014adam}
Kingma, D.P., Ba, J.: Adam: A method for stochastic optimization. arXiv
  preprint arXiv:1412.6980  (2014)

\bibitem{li2021neural}
Li, J., Chen, S., Cao, Y., Sun, Z.: A neural network approach to sampling based
  learning control for quantum system with uncertainty. Communications in
  Computational Physics  \textbf{30}(5),  1453--1473 (2021)

\bibitem{li2024bridging}
Li, P., Hao, J., Tang, H., Fu, X., Zhen, Y., Tang, K.: Bridging evolutionary
  algorithms and reinforcement learning: A comprehensive survey on hybrid
  algorithms. IEEE Transactions on evolutionary computation  (2024)

\bibitem{li2020fourier}
Li, Z., Kovachki, N., Azizzadenesheli, K., Liu, B., Bhattacharya, K., Stuart,
  A., Anandkumar, A.: Fourier neural operator for parametric partial
  differential equations. arXiv preprint arXiv:2010.08895  (2020)

\bibitem{li2024physics}
Li, Z., Zheng, H., Kovachki, N., Jin, D., Chen, H., Liu, B., Azizzadenesheli,
  K., Anandkumar, A.: Physics-informed neural operator for learning partial
  differential equations. ACM/IMS Journal of Data Science  \textbf{1}(3),
  1--27 (2024)

\bibitem{liu2021survey}
Liu, Y., Sun, Y., Xue, B., Zhang, M., Yen, G.G., Tan, K.C.: A survey on
  evolutionary neural architecture search. IEEE transactions on neural networks
  and learning systems  \textbf{34}(2),  550--570 (2021)

\bibitem{liu2024kan}
Liu, Z., Wang, Y., Vaidya, S., Ruehle, F., Halverson, J., Solja{\v{c}}i{\'c},
  M., Hou, T.Y., Tegmark, M.: Kan: Kolmogorov-arnold networks. arXiv preprint
  arXiv:2404.19756  (2024)

\bibitem{lu2021learning}
Lu, L., Jin, P., Pang, G., Zhang, Z., Karniadakis, G.E.: Learning nonlinear
  operators via deeponet based on the universal approximation theorem of
  operators. Nature machine intelligence  \textbf{3}(3),  218--229 (2021)

\bibitem{ong2010memetic}
Ong, Y.S., Lim, M.H., Chen, X.: Memetic computation—past, present \& future
  [research frontier]. IEEE Computational Intelligence Magazine  \textbf{5}(2),
   24--31 (2010)

\bibitem{ouyang2026learn}
Ouyang, K., Ke, Z., Fu, S., Liu, L., Zhao, P., Hu, D.: Learn from global
  correlations: Enhancing evolutionary algorithm via spectral gnn. In:
  Proceedings of the AAAI Conference on Artificial Intelligence. vol.~40, pp.
  24665--24673 (2026)

\bibitem{qin2005self}
Qin, A.K., Suganthan, P.N.: Self-adaptive differential evolution algorithm for
  numerical optimization. In: 2005 IEEE congress on evolutionary computation.
  vol.~2, pp. 1785--1791. IEEE (2005)

\bibitem{qiu2025evolution}
Qiu, X., Gan, Y., Hayes, C.F., Liang, Q., Xu, Y., Dailey, R., Meyerson, E.,
  Hodjat, B., Miikkulainen, R.: Evolution strategies at scale: Llm fine-tuning
  beyond reinforcement learning. arXiv preprint arXiv:2509.24372  (2025)

\bibitem{raissi2019physics}
Raissi, M., Perdikaris, P., Karniadakis, G.E.: Physics-informed neural
  networks: A deep learning framework for solving forward and inverse problems
  involving nonlinear partial differential equations. Journal of Computational
  physics  \textbf{378},  686--707 (2019)

\bibitem{stanley2019designing}
Stanley, K.O., Clune, J., Lehman, J., Miikkulainen, R.: Designing neural
  networks through neuroevolution. Nature Machine Intelligence  \textbf{1}(1),
  24--35 (2019)

\bibitem{storn1997differential}
Storn, R., Price, K.: Differential evolution--a simple and efficient heuristic
  for global optimization over continuous spaces. Journal of global
  optimization  \textbf{11}(4),  341--359 (1997)

\bibitem{tanabe2013success}
Tanabe, R., Fukunaga, A.: Success-history based parameter adaptation for
  differential evolution. In: 2013 IEEE congress on evolutionary computation.
  pp. 71--78. IEEE (2013)

\bibitem{tanabe2014improving}
Tanabe, R., Fukunaga, A.S.: Improving the search performance of shade using
  linear population size reduction. In: 2014 IEEE congress on evolutionary
  computation (CEC). pp. 1658--1665. IEEE (2014)

\bibitem{wang2021understanding}
Wang, S., Teng, Y., Perdikaris, P.: Understanding and mitigating gradient flow
  pathologies in physics-informed neural networks. SIAM Journal on Scientific
  Computing  \textbf{43}(5),  A3055--A3081 (2021)

\bibitem{wang2022and}
Wang, S., Yu, X., Perdikaris, P.: When and why pinns fail to train: A neural
  tangent kernel perspective. Journal of Computational Physics  \textbf{449},
  110768 (2022)

\bibitem{weber2003evolution}
Weber, B.H., Depew, D.J.: Evolution and learning: The Baldwin effect
  reconsidered. Mit Press (2003)

\bibitem{whitelam2021correspondence}
Whitelam, S., Selin, V., Park, S.W., Tamblyn, I.: Correspondence between
  neuroevolution and gradient descent. Nature communications  \textbf{12}(1),
  ~6317 (2021)

\bibitem{zhang2009jade}
Zhang, J., Sanderson, A.C.: Jade: adaptive differential evolution with optional
  external archive. IEEE Transactions on evolutionary computation
  \textbf{13}(5),  945--958 (2009)

\bibitem{zhang2007moea}
Zhang, Q., Li, H.: Moea/d: A multiobjective evolutionary algorithm based on
  decomposition. IEEE Transactions on evolutionary computation  \textbf{11}(6),
   712--731 (2007)

\bibitem{zhou2021survey}
Zhou, X., Qin, A.K., Gong, M., Tan, K.C.: A survey on evolutionary construction
  of deep neural networks. IEEE Transactions on Evolutionary Computation
  \textbf{25}(5),  894--912 (2021)

\bibitem{zitzler2001spea2}
Zitzler, E., Laumanns, M., Thiele, L.: Spea2: Improving the strength pareto
  evolutionary algorithm. TIK report  \textbf{103} (2001)

\end{thebibliography}

\end{document}